# Tip enhanced IR imaging with sub-10 nm resolution and hypersensitivity

Jian Li[1], Junghoon Jahng[2], Jie Pang[1], William Morrison[3], Jin Li[1], Eun Seong Lee[2], Jing-Juan Xu[1], Hong-Yuan Chen[1], Xing-Hua Xia[1]*.

[1]State Key lab of Analytical Chemistry for Life Science, School of Chemistry and Chemical Engineering, Nanjing University, 210023 Nanjing, China

[2]Center for nanocharacterization, Korea Research Institute of Standards and Science, Daejeon 34113, Republic of Korea

[3]Molecular Vista Inc., 6840 Via Del Oro, Suite 110, San Jose, CA 95119, USA.

*To whom correspondence should be addressed. E-mail: xhxia@nju.edu.cn (X.H.X)

**Abstract**: **IR spectroscopy has been widely used for chemical identification and quantitative analysis of reactions occurring in a specific time and space domains by measuring an average signal of the entire system[1]. Achieving IR measurements with nanometer-scale spatial resolution is highly desirable to obtain a detailed understanding of the composition, structure and function of interfaces[2-5]. The challenges in IR nanoscopy yet exist owing to the small molecular cross section and pristine optical diffraction limit. Although atomic force microscopy (AFM) based techniques, such as scattering-type scanning near-field optical microscopy and photothermal-induced resonance microscopy (PTIR), can acquire IR spectroscopy in a few tens of nanometer scale resolution[6-9], IR measurements with monolayer level sensitivity remains elusive and can only be realized under critical conditions[10,11]. Herein, we demonstrate sub-10 nm spatial resolution sampling a volume of ~360 molecules with a strong field enhancement at the sample-tip junction by implementing noble metal substrates (Au, Ag, Pt) in photo-induced force microscopy (PiFM). This technique shows versatility and robustness of PiFM, and is promising for application in interfacial studies with hypersensitivity and super spatial resolution.**

Photo-induced force microscopy (PiFM) is an extremely sensitive technique that transfers the optical absorption to the measurable force response on AFM, supporting the super resolution imaging and spectrum recording down to the AFM tip apex level. A schematic of the PiFM setup is shown in Figure 1. The commercial AFM microscope is equipped with an integrated off-axis parabolic mirror. A tunable mid-IR quantum cascade laser (QCL) is used as the excitation source, with a pulse width of 40 ns and an adjustable repetition rate, which is set to the difference frequency between the first and second mechanical resonances. PiFM detects the motion of AFM tip induced by the sample-tip interaction force. In this study, we use the second mechanical mode to detect the topography, operated in tapping mode and the first mechanical mode to detect the photoinduced force signal.

PiFM has different sample thickness dependence than photothermal-induced resonance microscopy (PTIR) [12]. However, PiFM is more sensitive to the molecules close to the tip, while

PTIR measures the expansion force from the bulk material far underneath the tip. Thus, PiFM is in principle more suitable to study interfacial properties with respect to tip-sample distance. A diagram of the forces involved in the PiFM measurement is shown in Figure 2a. In the mid-IR region, expansion-modulated van der Waals force has recently been proposed as the dominant term for molecules in PiFM operation[12]. The van der Waals force is proportional to the thermal expansion of sample under illumination, which can be written as,

$$\Delta \mathbf{F}_{vDW} \propto \Delta \mathbf{L} \propto \sigma \int |\mathbf{E}|^2 \mathbf{dz} \quad (1)$$

where, $\Delta \mathbf{L}$ is the tip-enhanced thermal expansion, $\sigma$ is the linear thermal expansion coefficient and E is the position dependent field strength.

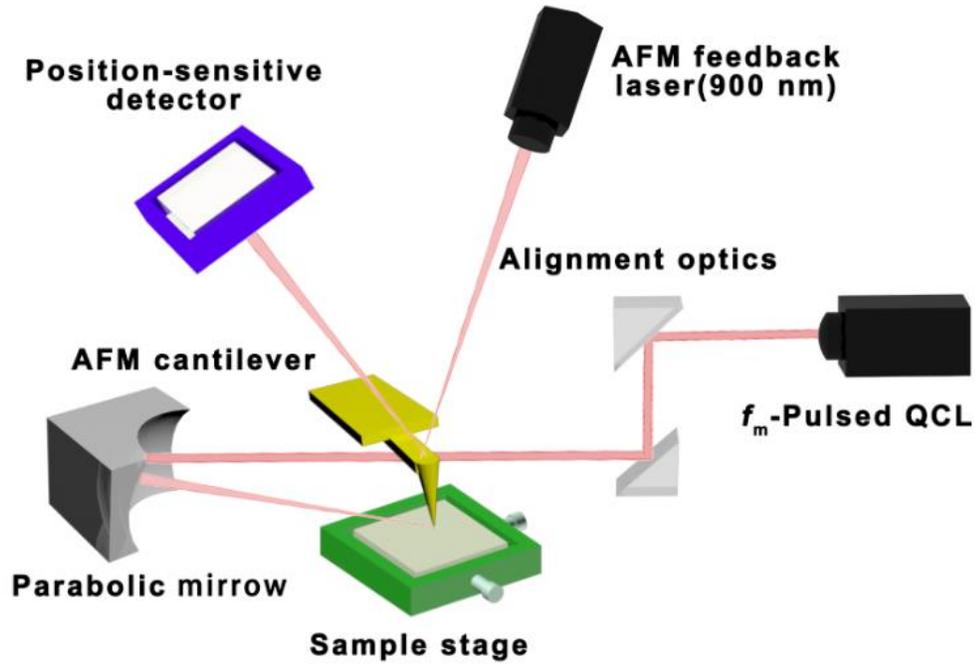

**Figure 1. Scheme of PiFM experiment.** Simultaneous record of topography and PiFM are performed with the first ($f_0$) and second ($f_1$) mechanical eigenmode resonances of the cantilever, respectively. The incident light is polarized along the tip axis and electrically triggered to pulse at fm = $f_1$ - $f_0$. A more in detailed mechanism of detection can be found in some previous works. [12,13]

The first demonstration of PiFM to measure thin films was pioneered by Derek et al[13] in 2016. They distinguished the patterns of 30 nm thick block copolymer (BCP) films and revealed the topography and composition relation within 10 nm spatial resolution. Thick samples were then performed and explored in organic solar cell [14, 15] and phonon polariton[16] studies. Expanding the PiFM measurements to a monolayer level will allow the exploration of important interfacial phenomena in heterogeneous samples, such as characterizing individual active sites of catalysts or identifying different reaction intermediates at catalytic surfaces[17,18]. However, interfacial study of monolayers using PiFM requires more critical experimental conditions than thick samples and until now has not been reported. Because monolayer measurements involve far fewer molecules than film measurements, the PiFM signal is expected to be extremely lower. Thus, an additional means of enhancing the PiFM signal is essential.

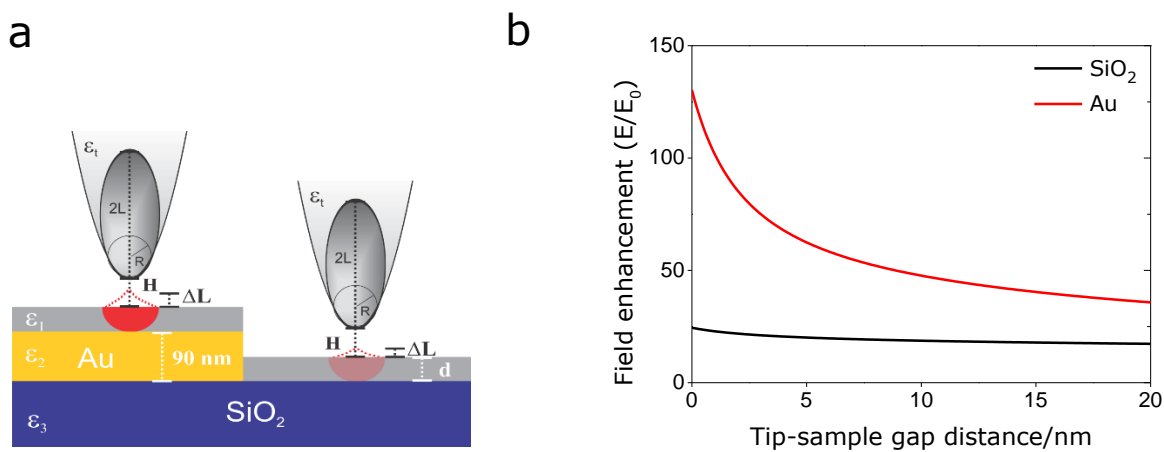

**Figure 2. Field distribution and PiFM on gold and SiO$_2$.**

**a,** Schematic for the tip enhanced PiFM set-up on gold surface. The tip is modeled as an ellipsoid with length of 2L. The H is the gap distance from the tip end to the sample surface. The ΔL is the tip-enhanced thermal expansion and $\varepsilon_1$, $\varepsilon_2$, $\varepsilon_3$, $\varepsilon_t$ are the permittivity of sample, gold film, SiO$_2$ and tip material, respectively.

**b,** Simulation of E field strength on the substrate surface with the distance between tip end and substrate surface. Blue curve: on SiO$_2$; Red curve: on gold. Both curves show highly distance dependence.

Herein, we demonstrate an electromagnetic enhanced approach to achieve PiFM measurements with monolayer level sensitivity. Since the PiFM signal is proportional to the square of the E field (formula 1), hypersensitive interfacial measurements on top of a metallic layer can be realized with the enhanced electromagnetic field at the tip-sample junction, as shown in Figure 2a. With the lightning-rod effect and multi-reflection process, electromagnetic field enhancement is generated in between the tip-gold gap and utilized to enhance the PiFM signal[19]. A simulation of E field strength on Au film with gap distance is plotted in Figure 2b. For comparison, the E field between a tip and SiO$_2$ is also plotted. Details of the simulation can be found in Supplementary section 2. With a small gap size, the E field in the tip-substrate gap is effectively enhanced on both substrates. On gold the E filed enhancement is more than 130, resulting in a 16900 times enhancement of PiFM signal, which is about 29 times larger than on $SiO_2$. Thus, a tip-gold set-up is more promising for monolayer detection and chosen for the monolayer detection.

Self-assembled monolayers (SAMs) of 1-mercaptohexadecanoic acid (MDHA) and 4-mercaptobenzoic acid (MBA) are prepared on gold films for the experimental demonstrations. The MDHA and MBA SAMs are fabricated on gold films with high packing density via the Au-S covalent bond. The formation of SAMs follows a standard procedure and is described in methods. Herein, MDHA is a linear molecule with long chain length, and the SAM thickness has been reported as 2.5 nm[20]; while MBA is a planar molecule with smaller size, and the SAM thickness has been reported as 0.9 nm[21].

**Table1. Calculated thermal expansion and van der Waals force of MDHA and MBA SAMs on gold and $SiO_2$ substrates.**

| On gold | MBA | MDHA |
|---|---|---|
| $\Delta L$ | 5 pm | 107 pm |
| $\Delta F_{vdW}$ | 1.6 pN | 13 pN |
| On SiO2 | MBA | MDHA |
| $\Delta L$ | 0.2 pm | 7 pm |
| $\Delta F_{vdW}$ | 0.07 pN | 0.9 pN |

The modulated van der Waals forces due to the thermal expansion of the two SAMs on gold and $SiO_2$ are calculated based on formula (1) respectively. The results are listed in Table 1 and the detailed calculation process can be found in the supplementary materials (Section 3). Since the noise level of PiFM is roughly ~ 0.1 pN, [12] MBA and MDHA are barely measurable on $SiO_2$ substrates but they are clearly measurable on Au films. Hence, for small molecules, our set-up is better than conventional PiFM detection in principle.

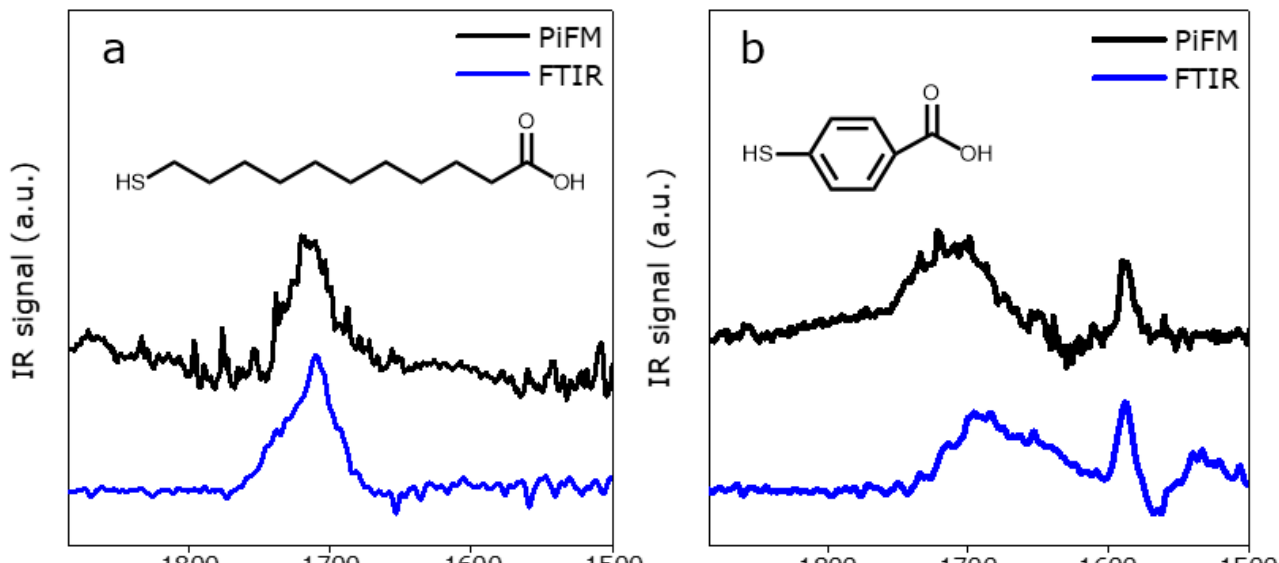

**Figure 3. PiFM and FTIR characterizations. a,** PiFM (black curve) spectrum of MDHA SAM on gold and ATR-FTIR (blue curve) spectrum of MDHA in aqueous; **b,** PiFM (black curve) spectrum of MBA SAM on gold and ATR-FTIR (blue curve) spectrum of MBA in aqueous.

Figures 3a and 3b show the PiFM spectra of MDHA and MBA SAMs and ATR-FTIR spectra for each molecule in solution, respectively. PiFM signal is recorded by taking the profile of the incident laser power as the reference. Generally, the characteristics of PiFM spectra are in good agreement with the macroscopic spectra. For MDHA, the absorption band at around 1710 cm$^{-1}$ corresponds to the carbonyl stretch C=O[22]. For MBA, two characteristic peaks are observed with PiFM. The peak at around 1700 cm$^{-1}$ corresponds to the carbonyl stretch C=O and the sharp peak at 1590 cm$^{-1}$ corresponds to the C=C aromatic stretch of the benzene ring [23]. These results demonstrate the feasibility of PiFM measurements on various molecular monolayers.

PiFM detection of MBA SAMs has been also successfully recorded on Pt and Ag films, which both possess high reflectance in the mid-IR (Supplementary Figure 1). This suggests the use of a tip-metal gap as a general method for monolayer level detection of PiFM, and also demonstrates that multiple substrate types can be used for this application. As is worthy to mention, since tips made of thermal evaporation technique have quite different signal sensitivity and are usually covered by containments, quantity relationship between the experiment and simulation is not established here.

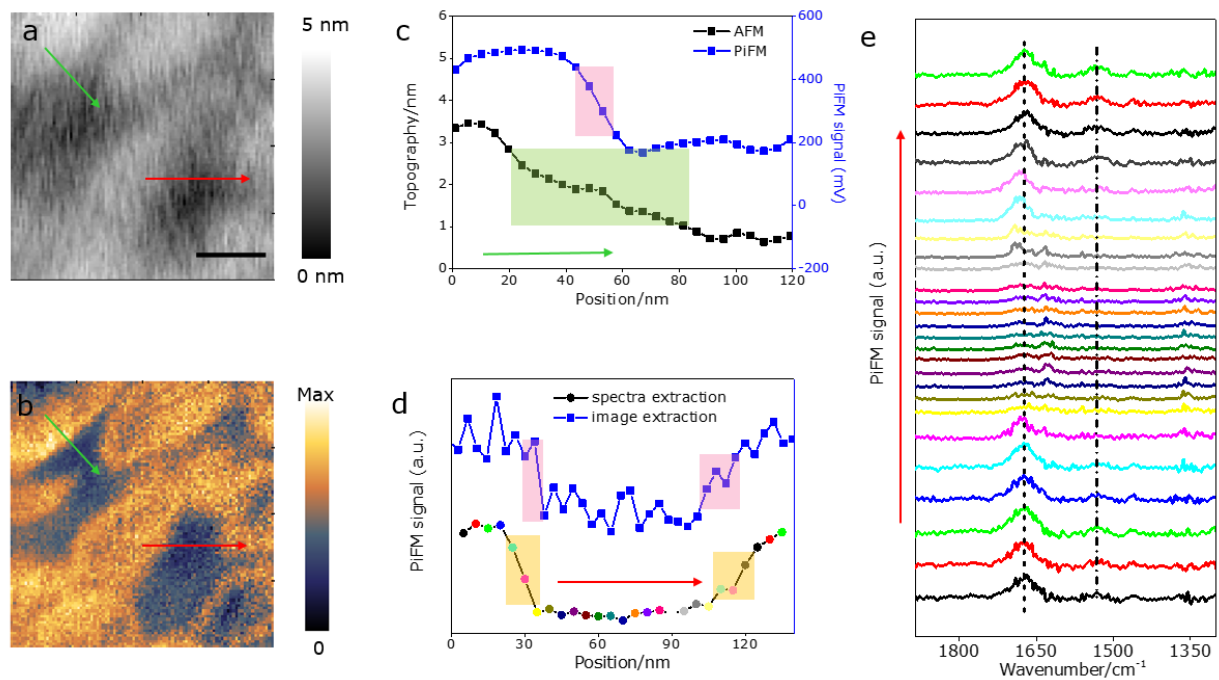

**Figure 4. PiFM record on BSA sub monolayer.**

    **a,** Topography of the disperse BSA on gold in tapping mode. Scale bar: 100 nm.

    **b,** PiFM image of the disperse BSA on gold recorded at 1660 cm$^{-1}$ at the same time with (**a**).

    **c,** Signal along the green arrow of topography (black curve) and PiFM image (blue curve).

    **d,** Extracted data point at 1660 cm$^{-1}$ from the PiFM spectra recorded at different points along the red arrow in (**b**) with 5 nm separation (colored curve) and the PiFM signal along the red arrow of PiFM image (blue curve).

    **e,** PiFM spectra recorded at the position colored-indicated in (**d**).

    The super-resolution imaging is then performed on a sparse distributed bull serum albumin (BSA) monolayer. BSA is a protein, and is selected as a demonstration of the feasibility of biological sample analysis. The fabrication of method is described in the method section. The adsorption of BSA (100 ng/mL) on gold substrate reaches equallium at around 30 min (Supplementary Figure 3a and 3b), thus a shorter time adsorption leads to a sub-monolayer level sample. Figure 4a shows a 375 nm×375 nm area of BSA sub-monolayer immobilized on gold film. The distribution of BSA is heterogeneous and the thickness is less than 5 nm, indicating the sub-monolayer nature of the sample since BSA is known to have a 14 × 4 × 4 nm size. Figure 4b is the PiFM image recorded at 1660 cm$^{-1}$ simultaneously with the topography. In general, AFM and PiFM images are consistent, with the PiFM showing much higher spatial resolution. Figure 4c shows the line traces from AFM and PiFM images along the green arrows. The point separation of the line trace is less than 5 nm. It's clear that the stage is obvious not sharp as read from the AFM image. The rectangle depicts the edge region and indicates that the spatial resolution of PiFM is less than 15 nm (could be smaller since the stage is a slope). Then, the

PiFM spectra are recorded along the red arrow with 5 nm separation. The strength at 1660 cm$^{-1}$ of each PiFM spectrum is extracted and color-plotted in Figure 4d with position, and the whole spectra are plotted in Figure 4e with the color corresponding to Figure 4d. The fluctuation of point extracted data shows great agreement with the same data extracted from the image, and the curve is more flat compared with the image extracted curve due to longer integration time (1 s for spectrum vs. 1/128 s for image). With the yellow rectangle region, the signal variation again suggests conservatively the resolution of the image is less than 10 nm. To our knowledge, this is the highest resolution for any reported tip-based IR technique on molecular monolayer level samples (both organic SAM and biological sample). We hypothesize that the PiFM resolution is so high because the strongest PiFM signal is generated within the E field hot spot, which is in principle smaller than the apex size of tip that determines the resolution of AFM. The hot spot size is further confirmed to be around 10 nm by a simulation on Comsol (Supplementary Figure 2). In addition, the PiFM point spectra offer more chemical information. The amide I band (at around 1660 cm$^{-1}$) and amide II band (at around 1530 cm$^{-1}$) are observed on the BSA layer, which well correspond with the FTIR measurements (Supplementary Figure 3c). In the middle lower part, no obvious IR characteristic peaks are observed, indicating no proteins deposited in this region, which is in good agreement with the AFM image. A great topography, PiFM and chemical composition relation is established with super resolution and hypersensitivity. In principle, if a tip with smaller radius are used, the resolution can be comparable to tip enhanced Raman spectroscopy used in ambient condition[18], and PiFM might be used to analyze interfacial samples with higher spatial resolution than ever studied in tip-based IR techniques.

The number of molecules contributing to the signal can be evaluated based on the observed PiFM spatial resolution. The surface density of MDHA SAM is estimated to be 4.5 molecules/nm$^2$ from the result of similar size 1-heptadecanethiol SAM on gold [24]. From the PiFM image, the effective detection region has a 10 nm diameter, leading the amount of detected molecules to be about 360. According to Table 1, the detected PiFM signal of MDHA SAM is even far larger than the noise level (0.1 pN, ref (12)), allowing space for the exploration of single molecule level detection. Due to the limitation of tip radius and substrate fabrication, single molecule level spatial resolution and sensitivity are not performed here but on the way. Furthermore, in our experiment, we did not utilized the largest power of laser, and additional improvement could thus be carried out by introducing more concentrated E field for example using a resonant tip like in s-SNOM[25]. An optical antenna could also be used as substrate to further focus the light. With these enhancements to the signal, it might be possible to reach the limit necessary for single molecule detection.

In summary, we have established a novel set-up capable of measuring monolayers with better than 10 nm spatial resolution, on a total volume of ~360 molecules. The method is demonstrated to be suitable for various SAMs on different metal substrates. It is expected that the tip enhanced PiFM can be used in modern interfacial studies to reveal composition and function at the nanoscale.

**Methods**

Methods and any associated references are available in the online version of the paper.

**Acknowledgements**

The authors acknowledge support from the National Key Research and Development Program of China (2017YFA0206500), National Key R&D Program of China (2017YFA0700500), the Natural National Science Foundation of China (21327902, 21635004). J. Jahng and E. S. Lee thank to Korea Research Fellowship Program through the National Research Foundation of Korea (NRF) funded by the Ministry of Science and ICT (2016H1D3A1938071) for the support. The authors thank P.K. Zhang, Y. Zhou and X.L. Ding for sample preparation.

**Author contributions**

X.H. Xia. supervised the project. J. Li and X.H. Xia conceived the ideas and performed the experiments. J. Jahng and E. S. Lee performed the electromagnetic field simulations and force calculations. J. Pang, M. Williams, J. Li, J.J. Xu and H.Y. Chen helped with experiments and analysis. All authors contributed to data interpretation and writing of the manuscript.

**Additional information**

Supplementary information is available in the online version of the paper. Reprints and permissions information is available online at www.nature.com/reprints. Correspondence and requests for materials should be addressed to X.H. Xia.

**Competing financial interests**

The authors declare no competing financial interests.